\documentclass[twocolumn,preprintnumbers,prl]{revtex4-2}
\usepackage{physics}
\usepackage{graphicx}
\usepackage{dcolumn}
\usepackage{bm}
\usepackage{color}

\begin{document}

\title{Observation and Manipulation of Optical Parametric Down-Conversion in the Langevin Regime}

\author{Yen-Ju Chen$^{1,2}$}
\email{These authors contributed equally to this work.}
\author{Chun-Yuan Cheng$^{1,2}$}
\email{These authors contributed equally to this work.}
\author{Tien-Dat Pham$^3$}
\author{Tzu-An Chen$^{1,2}$}
\author{Chang-Hau Kuo$^{1,2}$}
\author{Yen-Hung Chen$^{3,4,5}$}
\author{Chih-Sung Chuu$^{1,2}$}
\email{cschuu@phys.nthu.edu.tw}

\affiliation{$^1$Department of Physics and Center for Quantum Science and Technology, National Tsing Hua University, Hsinchu 30013, Taiwan}
\affiliation{$^2$National Center for Excellence in Quantum Information Science and Engineering, National Tsing Hua University, Hsinchu 30013, Taiwan}
\affiliation{$^3$Department of Optics and Photonics, National Central University, Jhongli 320, Taiwan}
\affiliation{$^4$Quantum Technology Center, College of Science, National Central University, Jhongli 320, Taiwan}
\affiliation{$^5$Center for Astronautical Physics and Engineering, National Central University, Jhongli 320, Taiwan}

\begin{abstract}

Quantum fluctuation plays a key role in the parametric down-conversion in the Langevin regime. In this paper, we report the experimental realization of optical parametric down-conversion in the Langevin regime on a chip. By precisely controlling the loss inherently tied to fluctuation, 
we observe the asymmetric Hong-Ou-Mandel dip---a hallmark of quantum fluctuation in the fluctuation-driven PDC, and the fluctuation-induced compression of single photons by nearly one order of magnitude. These findings pave the way for the manipulation of quantum fluctuation, quantum states, and system-reservoir interaction.

\end{abstract}


\maketitle


Parametric down-conversion (PDC) \cite{Harris1966} has been a cornerstone of quantum information processing, with applications ranging from quantum communication \cite{Ekert1991, Bennett1992-1, Bennett1992-2,Bennett1993,Bouwmeester1997} and quantum computing \cite{Shor1994,Knill2001,Kok2007,OBrien2007} to quantum metrology~\cite{Giovannetti2004}. In typical PDC \cite{Kwiat1995, Ou1988, Shih1988, Ou1999, Sensarn2009, Sensarn2010}, where the losses are negligible, the vacuum fields act as the drivers for the quantum fields in the Heisenberg picture \cite{Byer1968, Kleinman1968}. In the Langevin regime, where the losses are significant,
the vacuum field decays and the PDC is instead driven by quantum fluctuation \cite{Kolchin2007, Shwartz2012, Su2016}. However, previous studies of the fluctuation-driven PDC were possible only in the x-ray regime. For example, Shwartz \textit{et al.} \cite{Shwartz2012} demonstrated that the coincidence counts in the x-ray PDC align with the Langevin prediction. Although these works verified the applicability of the Langevin theory at x-ray wavelengths, the complexity of preparing and manipulating the synchrotron-based PDC sources poses a challenge for further or deeper studies. The realization of fluctuation-driven PDC in the optical regime, in which rich quantum-optical tools exist, will therefore allow exploration of its physics or applications more extensively. For example, will quantum fluctuation have any \textit{visible} effects on the quantum fields in the time or spatial domain? Can fluctuation-driven PDC be realized on a chip?

In this work, we experimentally realize the optical PDC in the Langevin regime on a waveguide chip. The precise control of the loss, inherently tied to quantum fluctuation, enables an in-depth study and manipulation of the quantum states generated from the fluctuation-driven PDC. In particular, we observe the hallmark of quantum fluctuation in the fluctuation-driven PDC--the asymmetric Hong-Ou-Mandel (HOM) dip, and the fluctuation-induced shaping or compression of single photons by nearly one order of magnitude. These experimental observations are in excellent agreement with the Langevin theory, which takes into account quantum fluctuations. Our work thus demonstrates not only the first optical PDC in the Langevin regime but also a novel way of manipulating single and entangled photons with Langevin noise. 


To describe the PDC process in a lossy medium, we use the Heisenberg-Langevin formalism \cite{Kolchin2007, Shwartz2012,Su2016}. The slowly varying envelope equations can be written as
\begin{equation}
\begin{aligned}
	& \frac{\partial a_{s}}{\partial z}+\left(\alpha_{s}+i\frac{\Delta k}{2}\right)a_{s}= i\kappa a^{\dag}_{i}+\sqrt{2\alpha_{s}}f_{s} \\
	& \frac{\partial a^{\dag}_{i}}{\partial z}+\left(\alpha_{i}-i\frac{\Delta k}{2}\right)a^{\dag}_{i}= -i\kappa^{*} a_{s}+\sqrt{2\alpha_{i}}f^{\dag}_{i}
\end{aligned}
\label{eq:coupledEq}
\end{equation}
Here, $a_{s}(\omega,z)$ and $a_{i}(\omega,z)$ are the annihilation operators of the signal and idler fields, respectively. The pump field is incident at $z=0$.  $\alpha_{s}$ and $\alpha_{i}$ are the absorption coefficients. $\Delta k(\omega)=k_{p}-k_{s}-k_{i}$ denotes the phase-mismatch. $\kappa$ is the nonlinear coupling coefficients. $f_{s}(\omega,z)$ and $f^{\dag}_{i}(\omega_{i},z)$ are quantum noise operators. By solving Eq.~(\ref{eq:coupledEq}), we are able to obtain the output operators $a_{s}(\omega,L)$ and $a^{\dag}_{i}(\omega_{i},L)$, where $L$ is the crystal length, with the commutators $[a_{j}(\omega_{1},z_{1}),a^{\dag}_{k}(\omega_{2},z_{2})]=(2\pi)^{-1}\delta_{jk}\delta(\omega_{1}-\omega_{2})\delta(z_{1}-z_{2})$ and $[f_{i}(\omega_{1},z_{1}),f^{\dag}_{i}(\omega_{2},z_{2})]=(2\pi)^{-1}\delta(\omega_{1}-\omega_{2})\delta(z_{1}-z_{2})$. The Glauber correlation function, representing the wave packets of the biphotons or heralded single photons, can then be calculated at $z=L$ to be 
\begin{equation}
\begin{aligned}
    G^{(2)}_{ab}(\tau)&=\langle a^{\dag}_{a}(t+\tau)a^{\dag}_{b}(t)a_{b}(t)a_{a}(t+\tau) \rangle\\
    &=(\frac{1}{2\pi})^{2}|\int_{-\infty}^{\infty}\phi_{ab}(\omega')e^{-i\omega'\tau}d\omega'|^{2}+R_{s}R_{i}\\
    &=|\phi_{ab}(\tau)|^{2}+R_{s}R_{i}
    \label{eq:G2}
\end{aligned}
\end{equation}
 where the subscript $ab$ is $si$ ($is$) if the idler (signal) photons are used as the trigger, $\tau$ is the time difference between the arrival times of the signal and idler photons at the detectors, and 
\begin{equation}
    \label{eq:phi}
    \begin{aligned}
      &\phi_{si}(\omega)=B^{*}(\omega)D(\omega)+\int^{L}_{0}dz'F^{*}(\omega,z')H(\omega,z')\\
      &\phi_{is}(\omega)=A^{*}(\omega)C(\omega)+\int^{L}_{0}dz'E^{*}(\omega,z')G(\omega,z')
    \end{aligned}
\end{equation}
are the biphoton wave functions in the frequency domain or biphoton spectra~\cite{Su2016}, consisting of parametric and Langevin terms. The single photon count rates of the signal and idler fields, $R_{s}=\langle a_{s}^{\dagger}(t,z)a_{s}(t,z)\rangle$ and $R_{i}=\langle a_{i}^{\dagger}(t,z)a_{i}(t,z)\rangle$, respectively, are
\begin{equation}
    \label{eq:RsRi}
    \begin{aligned}
        &R_{s}=\frac{1}{2\pi}\int_{-\infty}^{\infty} d \omega'  (|B(\omega')|^{2}+\int_{0}^{L}dz'|F(\omega',z')|^{2})\\
        &R_{i}=\frac{1}{2\pi}\int_{-\infty}^{\infty} d \omega'(|C(\omega')|^{2}+\int_{0}^{L}dz'|G(\omega',z')|^{2})
    \end{aligned}
\end{equation}
with the coefficients
\begin{equation}
    \label{eq:coefficient}
    \begin{aligned}
        \begin{bmatrix}A&B\\C&D\end{bmatrix}&=e^{-\textbf{M}L},\quad 
        \begin{bmatrix}E&F\\G&H\end{bmatrix}=e^{-\textbf{M}(L-z)}\begin{bmatrix}\sqrt{2\alpha_{s}}&0\\0&\sqrt{2\alpha_{i}}\end{bmatrix}, \\
        &\textbf{M}=\begin{bmatrix} \alpha_{s}+i\Delta k(\omega)/2 & -i\kappa \\ i\kappa^{*} & \alpha_{i}-i\Delta k(\omega)/2 \end{bmatrix}
    \end{aligned}
\end{equation}

\begin{figure}[t]
\centering
\includegraphics[width=9cm]{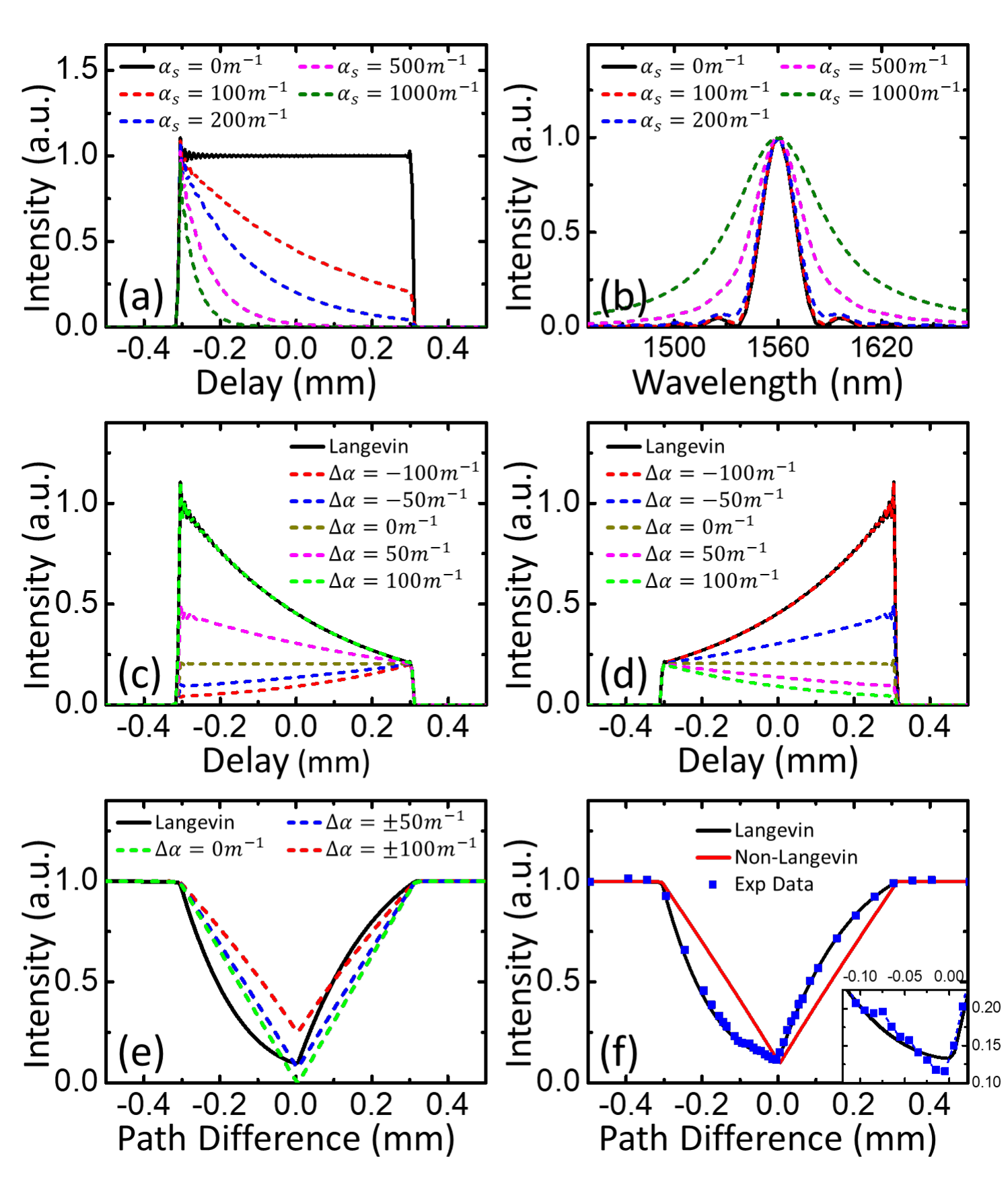}
\caption{\label{fig:FC}(a) Glauber correlation functions and (b) biphoton spectra for $\alpha_{s}=0, 100, 200, 500$ and $1000$~m$^{-1}$. (c) Idler-triggered and (d) signal-triggered Glauber correlation functions for $\alpha_{sum}=100$~m$^{-1}$. (e) The HOM interference as a function of path difference. The black solid line is the calculation with the Langevin terms, and the color lines (or dashed lines) are the results without the Langevin term for $\Delta\alpha =$ $-100$, $-50$, $0$, $50$, and $100$~m$^{-1}$. (f) The observed HOM interference (blue square) compared to theoretical predictions. Black and red curves are calculated by the Langevin and non-Langevin theories, respectively, with $\alpha_{s} = 87.00$~m$^{-1}$ and $\alpha_{i} = 29.77$~m$^{-1}$. The inset shows the region near the dip. All calculations consider a lossy region across the crystal length of $L =8$~mm.
}
\end{figure}


The effect of the fluctuation-tied loss on the quantum fields can be visualized by first examining the case where the idler field experiences loss while the signal field remains lossless, i.e. $\alpha_{s}>0$ and $\alpha_{i}=0$. Figure~\ref{fig:FC}(a) presents the Glauber correlation function of the biphotons for different magnitudes of $\alpha_s$. As $\alpha_s$ increases, the correlation function transitions from a rectangular profile to an exponentially decaying one. In the frequency domain, this corresponds to a transformation of the wave packet from a ${\rm sinc}^2$ shape to a Lorentzian-like profile, as depicted in Fig.~\ref{fig:FC}(b). Next, we consider the case where both $\alpha_{s}$ and $\alpha_{i}$ are nonzero. Figures ~\ref{fig:FC}(c) and \ref{fig:FC}(d) show the Glauber correlation functions for various differences $\Delta \alpha = \alpha_{s} - \alpha_{i}$ but with a constant sum $\alpha_{sum} = \alpha_{s} +\alpha_{i} = 100$~m$^{-1} $. Figures ~\ref{fig:FC}(c) and \ref{fig:FC}(d) (solid lines) display the idler- and signal- triggered biphoton waveforms, respectively, with the Langevin noise terms included. The idler- and signal- triggered waveforms exhibit an exponential decay and growth, respectively, multiplied by a unit box function, and are time reverses of each other. The calculation shows that the biphoton waveforms depend only on $\alpha_{sum}$ and are independent of $\Delta \alpha$. Consequently, the decay or growth constant of the wave packet is determined solely by $\alpha_{ sum}$. This is consistent with the fact that the biphoton waveform is a product of the photon number operators for both the signal and idler channels. 

In contrast, when the Langevin terms are excluded, the simulated waveforms [dashed lines in Fig.~\ref{fig:FC}(c) and Fig.~\ref{fig:FC}(d)] exhibit distinct dependencies on the individual loss coefficients. In this case, both the idler- and signal- triggered waveforms show exponential decay (or growth, depending on the sign of $\Delta \alpha$) with the same rate but different amplitudes. For example, the red dashed curves, corresponding to the minimum $\Delta \alpha$ with a negative sign ($\Delta \alpha = -100$~m$^{-1}$), show exponential growth for both waveforms. Although the signal-triggered waveform has a higher amplitude, consistent with the Langevin theory, the idler-triggered waveform exhibits a significant deviation, showing exponential growth instead of decay as the Langevin result. Similarly, the green dashed curves, representing the maximum $\Delta \alpha$ with a positive sign ($\Delta \alpha = 100$~m$^{-1}$), show both waveforms decaying exponentially at the same rate. However, while the idler-triggered waveform agrees with the Langevin theory, the signal-triggered waveform deviates. In the special case of $\Delta \alpha = 0$~m$^{-1}$ (brown dashed curves), both waveforms become identical squared functions, further separating from the prediction of the Langevin theory.

To explore the interplay between fluctuation and quantum fields, we exploit HOM interference in our experiments. When two indistinguishable photons enter the different input ports of a 50:50 beam splitter, they exit at the same output port because of bosonic commutation relations, resulting in zero coincidence counts between the detectors at different output ports. The Glauber correlation function in this scenario is given by
\begin{equation}
    \label{eq:HOMg2}
    \begin{aligned}
        G^{(2)}(\tau)=|\frac{1}{2}[\phi_{si}(\Delta t + \tau )-\phi_{is}(\Delta t - \tau)]|^{2}
    \end{aligned}
\end{equation}
where $\Delta t$ is the time difference between the two photons arriving at the beam splitter. Fig. ~\ref{fig:FC}(e) shows the calculated HOM interference. The Langevin theory predicts an identical and asymmetric interference pattern for different $\Delta \alpha$ (solid black line). 
Specifically, the second derivative of the interference pattern is positive for negative delays and negative for positive delays, which arises from the asymmetry of the wave packet due to loss inherently tied to fluctuation. In contrast, the non-Langevin theory predicts a symmetric pattern around the zero delay, as shown by the dashed lines in Fig.~\ref{fig:FC}(e). In addition, the waveform and visibility also vary for different $\Delta \alpha$. The visibility reaches the maximum of 1 when $\Delta \alpha = 0$ and decreases as $\Delta \alpha$ deviates from zero.

\begin{figure}[t]
\centering
\includegraphics[width=8cm]{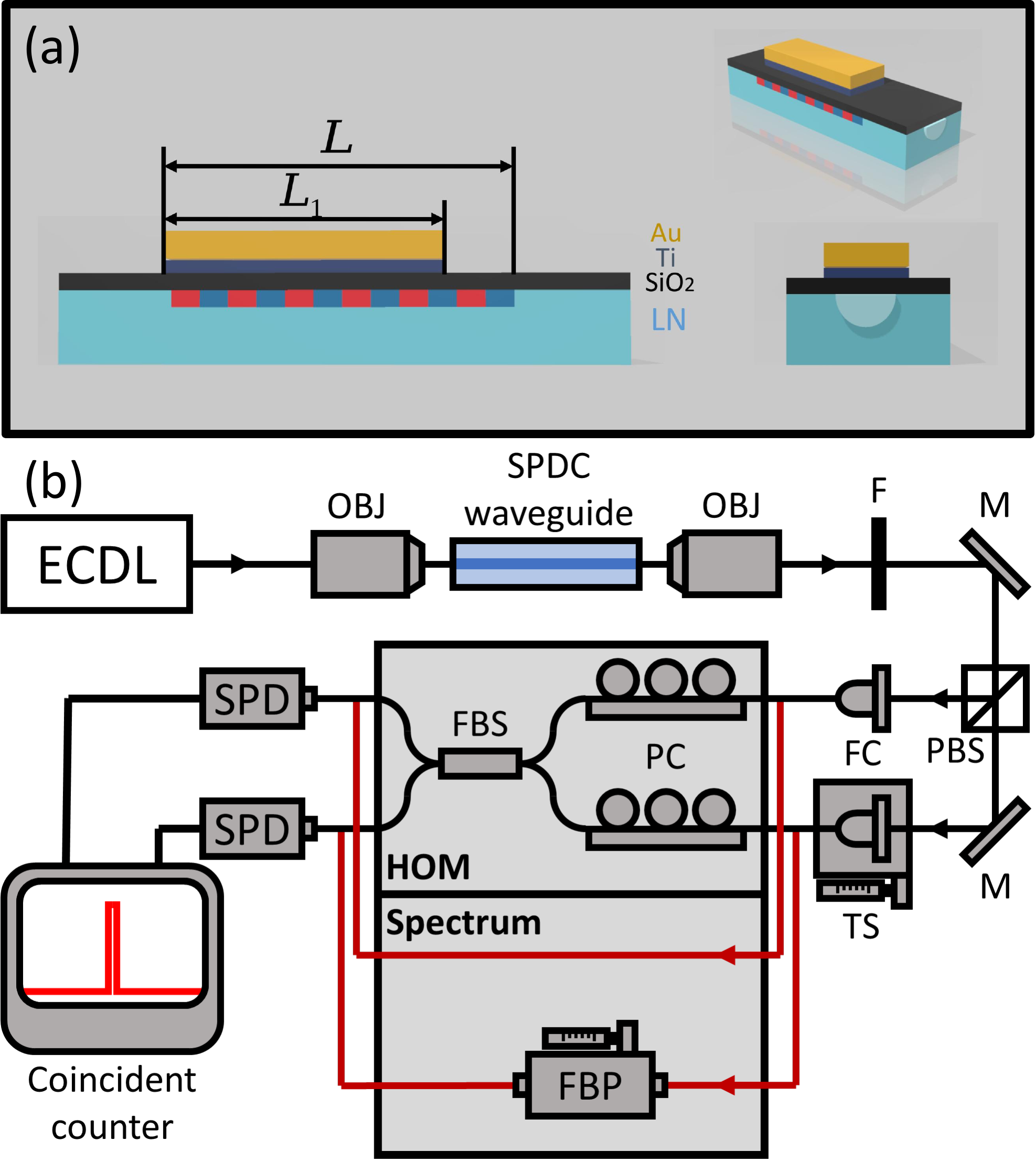}
\caption{\label{fig:setup} (a) Schematic of the Ti-diffused PPLN waveguide sample. The multilayer structure above the PPLN waveguide consists of 50 nm thick SiO$_{2}$ and 15 nm thick Ti layers, followed by a 200 nm layer of Au. Different lengths of the Ti/Au layers $L_{1}$ are fabricated. (b) The experimental setup of HOM interference (black optical paths) and spectrum measurement  (red optical paths). ECDL: external-cavity diode laser, OBJ: objective, F: long-pass filter, M: mirror, PBS: polarization beam splitter, FC: fiber coupler, TS: translation stage, PC: polarization controller, FBS: fiber beam splitter, FBP: fiber band-pass filter, SPD: single photon detector.
}
\end{figure}

Our experimental setup is illustrated in Fig. \ref{fig:setup}. To prepare the PDC source with controllable inherent loss, we utilize a nonlinear crystal waveguide pumped by a 780 nm external cavity laser diode, which is focused using a 10x microscope objective. The structure of our sample is illustrated in Fig.~\ref{fig:setup}(a). The waveguide consists of a 3 cm long lithium niobate substrate with a Ti-diffused periodically poled lithium niobate (PPLN) section \cite{Burns1979, Tanzilli2001}. The poling region is 8 mm long with a poling period of 9.6 $\mu$m, optimized for the type-II quasi-phase matching. The generated signal and idler photons are spatially separated by a polarization beam splitter and then coupled into single-mode fibers for HOM interference. The inherent loss is introduced by depositing an absorption layer on the waveguide, separated by a buffer layer. The loss can be tuned by adjusting the thickness of the buffer layer. The patterning of the metal layers can also be tailored to implement various system-reservoir interactions. In our work, a 115 nm dielectric buffer layer of SiO$_{2}$ is deposited on the waveguide, followed by a 15 nm Ti layer and a 200 nm Au layer. The resulting absorption coefficients, as predicted by finite-difference time-domain (FDTD) simulations, are approximately 63.97~m$^{-1}$ for the TM mode (signal photons) and 21.89~m$^{-1}$ for the TE mode (idler photons).

\begin{figure}[t]
\centering
\includegraphics[width=8.5cm]{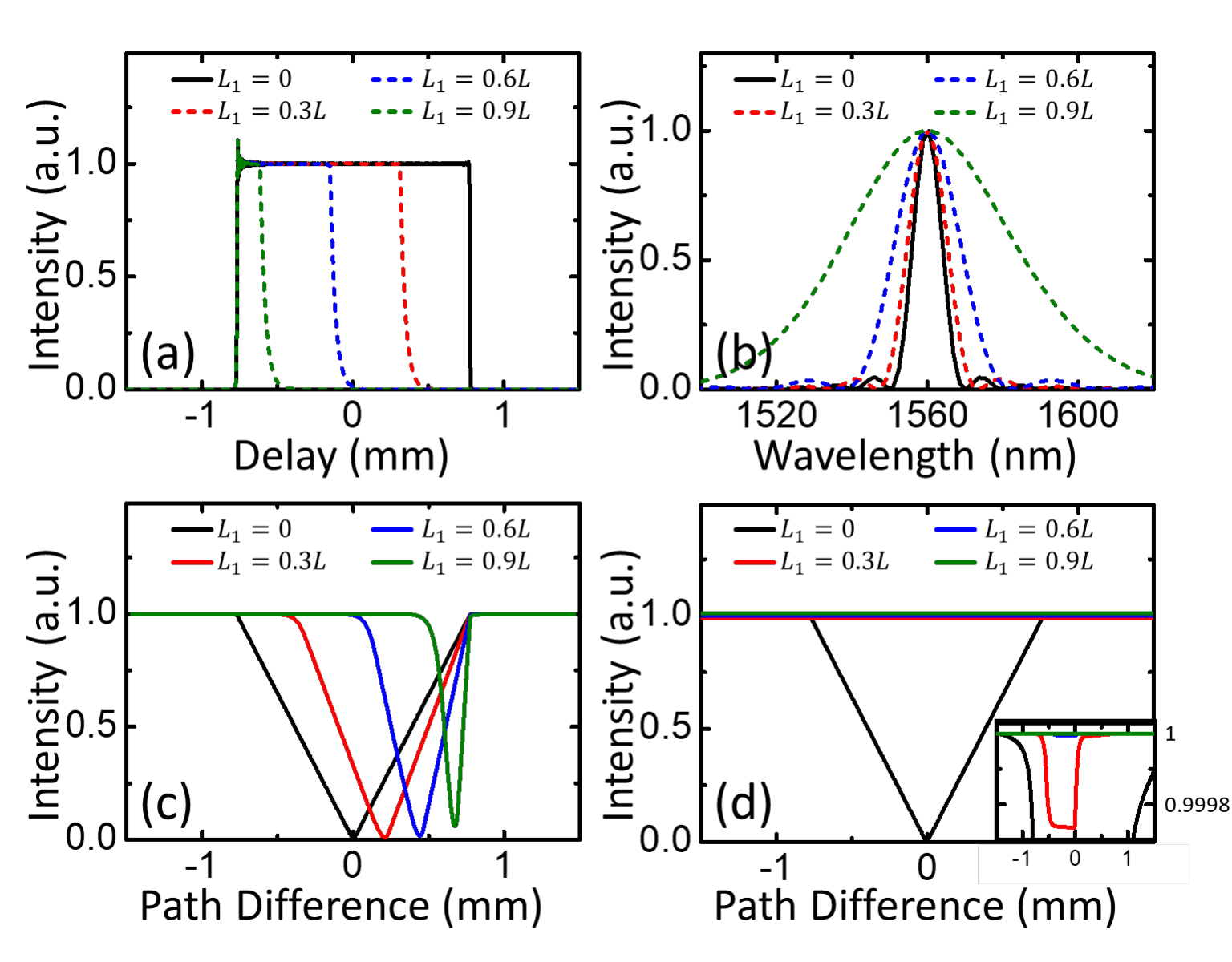}
\caption{\label{fig:PC_sim}
Fluctuation-driven PDC with position-dependent loss for $L=2$~cm, $\alpha_{s1}=1100$~m$^{-1}$, $\alpha_{i1}=60$~m$^{-1}$, and $\alpha_{i2}=\alpha_{s2}=0$~m$^{-1}$. (a) Glauber correlation functions. (b) Biphoton spectra. (c) HOM interferences. The FWHM widths and visibilities are 0.78 mm (1), 0.60 mm (0.981), 0.37 mm (0.967), and 0.14 mm (0.890) for $L_{1} = 0$, $0.3 L$, $0.6 L$, and $0.9 L$, respectively. (d) HOM interferences without the Langevin term.
}
\end{figure}

Figure \ref{fig:FC}(f) shows the HOM interference (blue squares) between the signal and the idler photons, with the absorption layer (the lossy region) covering the entire poling region of the waveguide. The experimental data are in good agreement with the Langevin theory (black line), which predicts an asymmetric pattern, and clearly deviates from the non-Langevin theory (red line). The interference visibility of 0.793 is also close to the visibility 0.773 predicted by the Langevin theory. The small deviation near the dip, as shown in the inset, is likely a result of defects in the Au film, which alter the absorption coefficient and cause waveform distortion. The actual buffer layer thickness may also be slightly thinner than the designed 115 nm, together with the non-negligible loss in the waveguide ($\sim 7$~m$^{-1}$), leading to a higher $\alpha_{sum}$ than that estimated by the FDTD simulation. 


The fluctuation-induced shaping of single photons suggests that the quantum fields may be further manipulated by controlling the loss. To explore such a possibility, we consider $\alpha_{i}=\alpha_{i1}$ in the region $0\leq z < L_{1}$ and $\alpha_{i}=\alpha_{i2}$ in the region $L_{1}\leq z \leq L$, while $\alpha_{s} = 0$ throughout the crystal. Figures \ref{fig:PC_sim}(a) and \ref{fig:PC_sim}(b) show the calculated Glauber correlation functions and biphoton spectra, respectively. The resulting wave packet has a rectangular shape with a temporal length of $L-L_{1}$ followed by an exponential decay. As $L_{1}$ increases, the wave packet (spectrum) becomes shorter (broader), providing a novel way of generating ultrashort single photons. Figures \ref{fig:PC_sim}(c) and \ref{fig:PC_sim}(d) show the calculated HOM profiles with and without Langevin terms, respectively. When Langevin noise is present, the coincidence counts exhibit a HOM dip with a FWHM width corresponding to the coherence length of the wavepacket in Fig. \ref{fig:PC_sim}(a). The visibility of the HOM dip decreases with $L_{1}$ as a consequence of the increasing portion of the exponential decay induced by the loss, which also leads to the asymmetry of the HOM dip.
In contrast, if Langevin terms are excluded, there is almost no interference dip, showing the importance of including Langevin noise in describing the HOM interference.

\begin{figure}[t!]
\centering
\includegraphics[width=8.5cm]{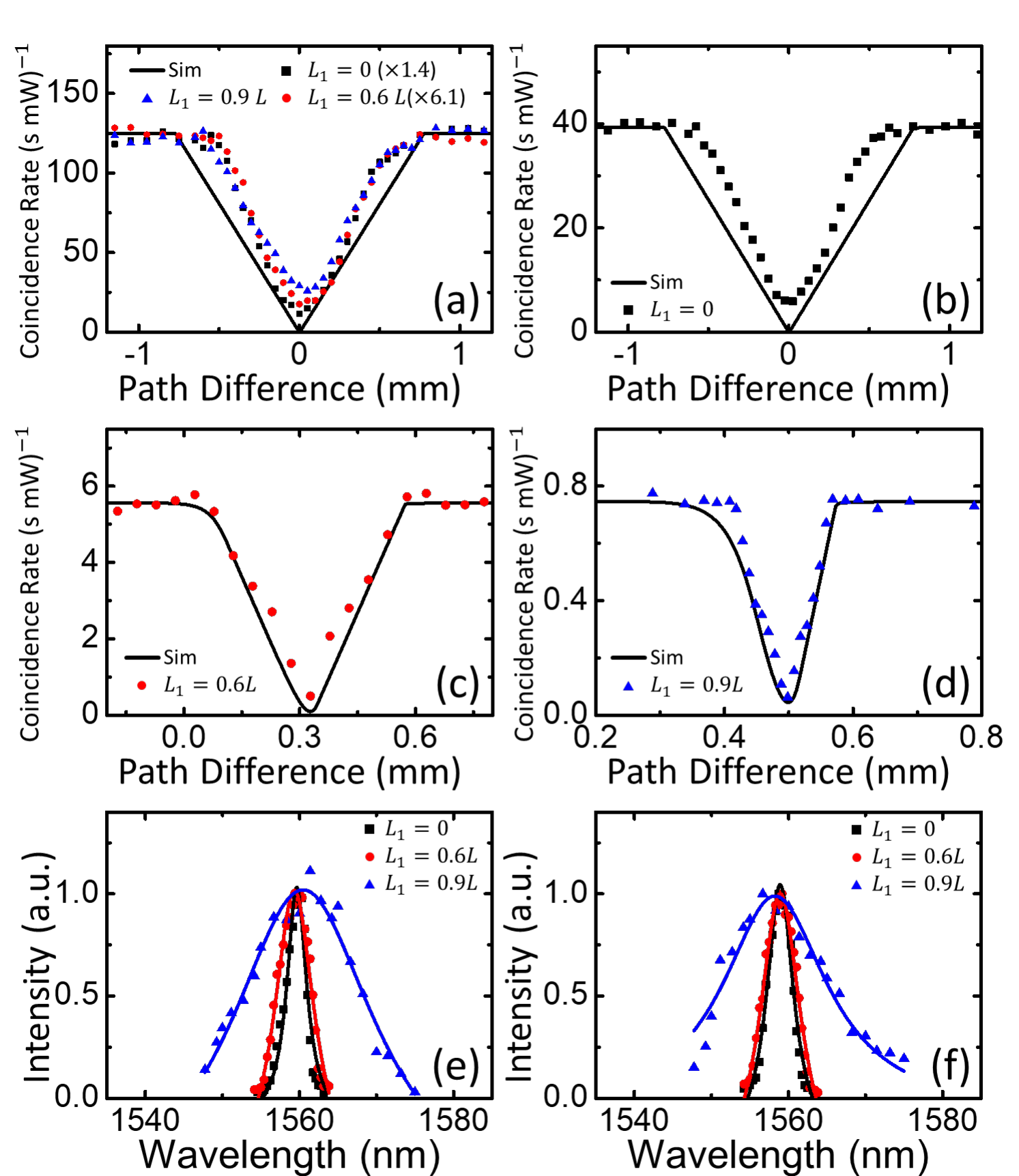}
\caption{\label{fig:PC_Data}
Manipulation of optical PDC in the Langevin regime. (a) Measured HOM dips without the SiO$_2$ and Ti/Au layers. The coincidence rates per mW of pump power for $L_1=0$ and $0.6L$ are multiplied by 1.4 and 6.1, respectively. (b) Measured HOM dip with the SiO$_2$ layer. (c) Measured HOM dip with the SiO$_2$ and Ti/Au layers ($L_1=0.6 L$). (d) Measured HOM dip with the SiO$_2$ and Ti/Au layers ($L_1=0.9 L$). The visibilities are 0.746, 0.840, and 0.851 in (b), (c), and (d) at a pump power (before the objective) of 0.3, 4, and 4 mW, respectively. The black curves are the corresponding theoretical predictions in Fig.~\ref{fig:PC_sim}(c). (e) Measured spectra of the heralded idler photons. (f) Measured spectra of the heralded signal photons.
}
\end{figure}

To demonstrate the manipulation of the fluctuation-driven PDC, we fabricate samples with a 3 cm long lithium niobate substrate and a $L=2$ cm long poling region. The poling region is partially covered by an absorption layer where a 15 nm Ti layer and a 200 nm Au layer are deposited with two different lengths: $L_{1} = 0.6 L$ and $0.9L$. A thinner buffer layer of 50 nm SiO$_{2}$, along with the Ti/Au layer, is used to enhance the effect of the fluctuation-tied loss, resulting in absorption coefficients of approximately $1100$~m$^{-1}$ for the TM mode (signal photons) and $60$~m$^{-1}$ for the TE mode (idler photons). The observed HOM interferences before and after the SiO$_{2}$ or Ti/Au layers are deposited are shown in Fig.~\ref{fig:PC_Data}. Before introducing the inherent loss (or the SiO$_{2}$ and Ti/Au layers) [Fig.~\ref{fig:PC_Data}(a)], the length of the wave packets inferred from the FWHM width of the HOM dip is approximately 0.61 mm (data points), slightly shorter than the theoretical prediction of 0.65 mm (black curve). The subsequent deposition of the SiO$_2$ layer [Fig.~\ref{fig:PC_Data}(b)] does not show a significant change in the HOM interference [as compared to the black squares or $L_{1}=0$ in Fig.~\ref{fig:PC_Data}(a)]. However, after the Ti/Au layer is deposited, the widths of the HOM dip and wave packet show an observable reduction, a phenomenon which we term single-photon tapering. For the sample with $L_{1}=0.6 L$, the corresponding length of the wave packet is 0.24 mm as shown in Fig.~\ref{fig:PC_Data}(c), which is 2.54 times shorter than before introducing the inherent loss. The length of the wave packet is further reduced to 92 $\mu$m or by 6.7 times in Fig.~\ref{fig:PC_Data}(d) with a SiO$_{2}$/Ti/Au layer for $L_{1}=0.9 L$. Although the pair rate decreases with increasing absorption layer coverage, the observed HOM interferences are in good agreement with theoretical predictions (black curves), showing a reduction in the width of wave packets. The discrepancy may arise from deviations in the deposited absorption layers or from the fabricated poling structures and waveguides.


We further investigate the effect of the Langevin noise on the spectral properties of signal and idler photons using
the experimental setup in Fig.~\ref{fig:setup}(b). By passing the heralded signal or idler photons through a tunable band-pass filter (1-nm bandwidth), the spectrum (approximately proportional to $|\phi_{si}(\omega)|^2$ or $|\phi_{is}(\omega)|^2$) is measured by scanning the filter's center wavelength. 
The idler spectra obtained are shown in Fig.~\ref{fig:PC_Data}(e) and are broadened in the presence of Langevin noise. The FWHM  bandwidths are 2.89, 4.87, and 16.6 nm for samples with $L_{1} = 0$, $0.6 L$ and $0.9 L$, respectively. Similarly, the signal spectrum shows significant spectral broadening, with FWHM bandwidths of 3.53, 4.95, and 13.35~nm for $L_{1} = 0$, $0.6 L$ and $0.9 L$, respectively. 
Despite a notable difference in channel loss between the signal and idler photons, 
the nonlocality of time-energy-entangled photons ensures that their spectra exhibit a similar broadening or $|\phi_{si}(\omega)|^2 = |\phi_{is}(\omega)|^2$ in a way analogous to nonlocal dispersion cancellation and modulation~\cite{Franson1992,Harris2008}.



In summary, we demonstrate the first optical PDC in the Langevin regime. The rich quantum-optics tools and techniques in the optical regime make possible the realization and manipulation of the fluctuation-driven PDC on a waveguide chip. 
Through HOM interference, we observe the asymmetry and single-photon tapering inherently tied to quantum fluctuation. By precisely controlling the loss inherently tied to fluctuation, we compress the single-photon wave packet to 11.5\% of its initial size, with a length approximately 100 times the wavelength.
Compression of the wave packet further down to a single cycle may be achieved by increasing the absorption layer coverage, reducing the buffer layer thickness, replacing the Au layer to obtain a larger absorption coefficient in the near-infrared range, or embedding (or doping) absorptive material into the waveguide. 
Compared with other techniques such as the use of very short nonlinear crystals, chirped quasi-phase-matching, or intensity modulators \cite{Harris2007,Balic2005,Kolchin2008,Chuu2012,Wu2019}, which also introduces a reduction in the pair generation rates, our approach does not require sophisticated fabrication of the poling structure or nonlinear crystals with a length on the order of the photon wavelength. We note that while Langevin noise is considered in the theory of cavity-based squeezed light generation \cite{Gatti1997}, the Langevin noise studied here is relevant in a single-pass parametric down-conversion at the single-photon level. Our findings thus pave the way for the manipulation of quantum fluctuation, quantum states, and system-reservoir interaction.

\textit{Acknowledgments.}---This work was supported by the National Science and Technology Council of Taiwan (113-2119-M-007-012).



\end{document}